# Autocomplete 3D Sculpting

Mengqi Peng    Jun Xing    Li-Yi Wei

The University of Hong Kong

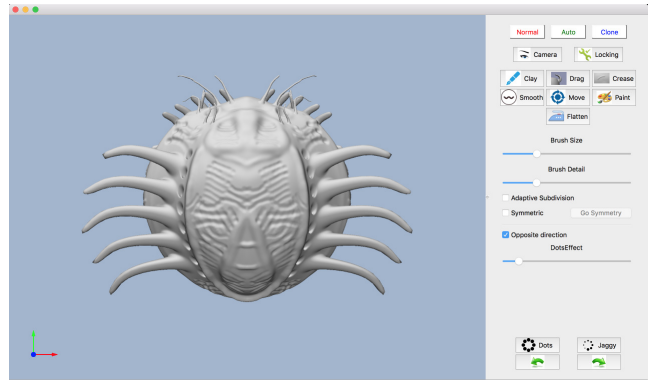

**Figure 1:** *User interface of our system. The interface consists of a sculpting canvas (left) and a widget panel (right). The widget panel provides the usual sculpting tools, brush parameters such as size, and mode controls unique to our autocomplete system.*

## Abstract

Digital sculpting is a popular means to create 3D models but remains a challenging task for many users. This can be alleviated by recent advances in data-driven and procedural modeling, albeit bounded by the underlying data and procedures.

We propose a 3D sculpting system that assists users in freely creating models without predefined scope. With a brushing interface similar to common sculpting tools, our system silently records and analyzes users' workflows, and predicts what they might or should do in the future to reduce input labor or enhance output quality. Users can accept, ignore, or modify the suggestions and thus maintain full control and individual style. They can also explicitly select and clone past workflows over output model regions. Our key idea is to consider *how* a model is authored via dynamic workflows in addition to *what* it is shaped in static geometry, for more accurate analysis of user intentions and more general synthesis of shape structures. The workflows contain potential repetitions for analysis and synthesis, including user inputs (e.g. pen strokes on a pressure sensing tablet), model outputs (e.g. extrusions on an object surface), and camera viewpoints. We evaluate our method via user feedbacks and authored models.

**Concepts:** •**Human-centered computing** → User interface design; •**Computing methodologies** → Shape modeling;

## 1 Introduction

3D modeling is ubiquitous in various applications, but demands significant expertise and efforts to create outputs with sufficient quality and complexity. This is particularly so for models consisting of repetitive structures and details, as common in many natural and man-made objects.

To help author 3D models, significant research has been devoted to methods based on data [Funkhouser et al. 2004; Mitra et al. 2013] or procedures [Emilien et al. 2015; Nishida et al. 2016]. These methods mainly focus on *what* the model is shaped instead of *how* it is authored, and the output scope is delineated by the underlying data or procedures. The processes of how models are authored or created by users, termed workflows, contain rich information that can facilitate a variety of modeling tasks based on individual user styles [Denning and Pellacini 2013; Chen et al. 2014; Denning et al. 2015; Salvati et al. 2015]. However, it remains unclear whether and how such workflows can help users create 3D models, especially under interactive interfaces such as digital sculpting, a popular means to author organic shapes with individual styles.

We propose a 3D sculpting system that assists users in freely creating models without pre-existing data or procedures. With a common brushing interface, our system analyzes what users have done in the past and predicts what they might or should do in the future, to reduce input workload and enhance output quality. The predictions are visualized as suggestions over the output model without disrupting user practices. Users can choose to accept, ignore, or modify the suggestions and thus maintain full control. They can also select prior workflows from the model and clone over other regions. The rich information contained in the workflows allows our method to outperform prior methods based on geometry. Similar to existing sculpting tools, our interface provides *surface* brushes for local details such as displacements and *freeform* brushes for large scale changes such as extrusions. Our system is intended for users with varying levels of expertise and models of different types without requiring pre-existing geometry data or procedural rules. In addition to sculpting strokes, our method also considers camera movements, which are often repetitive, predictable, and correlate well with the underlying shapes and brush strokes [Chen et al. 2014].

Inspired by recent works on predictive user interfaces that analyze workflows to assist 2D drawings [Xing et al. 2014] and animations [Xing et al. 2015], our premise is that 3D sculpting often consists of repetitive user actions and shape structures, and thus both the input process and output outcome can be predictable.

However, unlike 2D drawings and animations in which the base domain is a simple planar canvas, for 3D sculpting the base domain is a 3D object. Thus, even though surface brushes place local displacements analogous to paint brushes add local colors, they can reside over surface regions with different orientations and curvatures. Furthermore, unlike the planar canvas which remains invariant during 2D sketching, the 3D shape can undergo large scale changes by freeform brushes. Such changes might not even be functions over the domain surfaces and thus completely beyond the methodologies in [Xing et al. 2014; Xing et al. 2015]. Our key idea is to factor out the contextual parts of workflow positions via proper local parameterizations during analysis (e.g. repetition detection, clone select) and factor back the contexts during synthesis (e.g. suggestion, clone paste) while keeping track of geometry signatures such as surface normal and curvature all the time. For more holistic analysis and prediction, we combine different aspects of the workflows, including 2D inputs (e.g. pen strokes on a pressure tablet), 3D outputs (e.g. brushes on an object surface), and camera movements.

We conduct a pilot user study to show that our system can help users on both objective performance and subjective satisfaction for a variety of output models. In sum, the main contributions of this paper include:

- The idea that dynamic workflows can augment static geometry for better interactive 3D modeling;
- An autocomplete user interface for more friendly and effective 3D sculpting via hint/suggestion, workflow clone, camera control, and other features;
- Methods that analyze what users have done in the past to predict what they may and/or should do in the future, based on a similarity measure considering local frame and geometry, and combinations of 2D inputs (e.g. on a tablet), 3D outputs (e.g. extrusions on an object surface), and camera viewpoints.

## 2 Previous Work

**Data-driven and procedural modeling** Creating models from scratch is challenging, but similar objects or parts often already exist. Analyzing existing model geometry for novel synthesis has been a very active area of research [Funkhouser et al. 2004; Mitra et al. 2013] encompassing a variety of topics, such as suggestion [Chaudhuri and Koltun 2010], repetition [Bokeloh et al. 2011], symmetry [Mitra et al. 2006], style [Lun et al. 2016], fabrication [Schulz et al. 2014], and functional interaction [Hu et al. 2016]. Common model structures can also be abstracted into procedural rules for offline or interactive modeling [Ebert et al. 2002; Emilien et al. 2015; Nishida et al. 2016].

The outputs of these methods are naturally limited by the scopes of the underlying data and procedures. Our method, in contrast, aims to assist users to explore and create models [Cohen-Or and Zhang 2016] in their individual styles and preferences.

**Modeling interface** Popular modeling interfaces have been designed for mechanic models (e.g. AutoCAD) and organic shapes (e.g. ZBrush). A variety of enhancements have also been proposed for stability [Umetani et al. 2012], printability [Zehnder et al. 2016], reconfigurability [Garg et al. 2016], tiling [Guérin et al. 2016], and collaboration [Talton et al. 2009].

Following the line of suggestive interfaces for 3D drawing and sketching [Igarashi and Hughes 2001; Tsang et al. 2004], our system aims for a traditional sculpting interface with enhancement in analyzing, predicting, and suggesting workflows.

**Workflow-assisted authoring** Workflows [Nancel and Cockburn 2014] have been investigated for various authoring tasks in 2D image editing [Chen et al. 2011; Chen et al. 2016], sketching [Xing et al. 2014], and animation [Xing et al. 2015], as well as in 3D modeling such as visualization [Denning et al. 2011; Denning et al. 2015], revision control [Denning and Pellacini 2013], view selection [Chen et al. 2014], and collaboration [Salvati et al. 2015]. These workflows can be recorded during authoring, or inferred a posteriori [Fu et al. 2011; Hu et al. 2013; Tan et al. 2015].

As reported in [Santoni et al. 2016], even for complex objects, the workflows often consist of predictable brush choices and operations. Our method analyzes 3D sculpting workflows to autocomplete potential repetitions.

## 3 User Interface

The user interface (Figure 1) of our prototype sculpting system follows the brush model as in popular digital sculpting tools such as Blender [Blender Foundation 2016], Sculptris [Pixologic 2011], Zbrush [Pixologic 2015], MeshMixer [Schmidt and Singh 2010], etc. Our system supports *surface* brushes for local details via displacement, and *freeform* brushes for large scale change such as extrusion. Users can sculpt as usual while our system silently records and analyzes the sculpting workflows. All brush operations can be combined with the main functions: hint, workflow clone, camera control, as well as other features such as workflow lock.

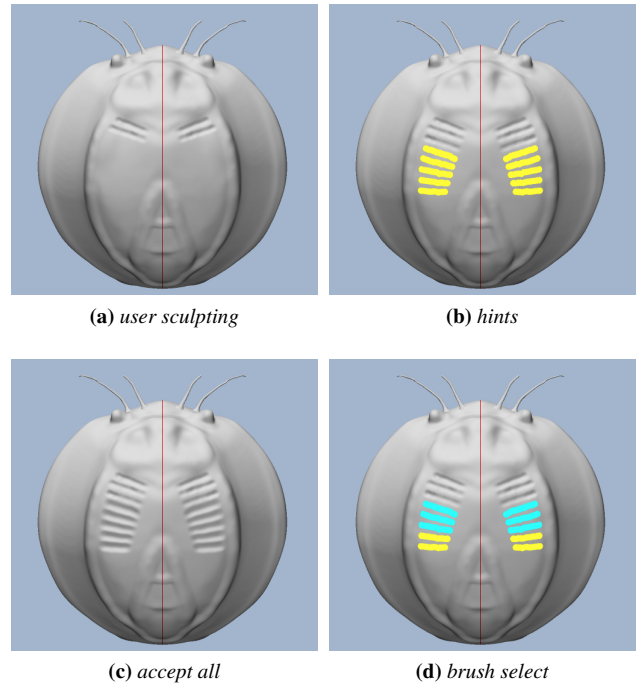

**(a)** *user sculpting*  **(b)** *hints*

**(c)** *accept all*  **(d)** *brush select*

**Figure 2:** *Hint example. During user sculpting (a), our system suggests potential future repetitions as displayed in transparent yellow (b). Users can ignore the hint and continue sculpting, accept all hints via a hot key (c), or partially select a subset of the hints (d) as shown in transparent blue.*

### 3.1 Hint

Our system automatically analyzes users' sculpting workflows on the fly and predicts what they might or should sculpt in the near future. These predictions are suggested as hints on our user interface. Figure 2 shows an example where the user is sculpting detailed features on an object. As more brushes are added, our system automatically analyzes the past strokes and predicts what brushes the user might want to perform next, as shown in Figure 2b. The suggestions are shown transparently over the object surface. Users can ignore the suggestions and continue sculpting as usual, accept all the suggestions via a hot key (Figure 2c), or partially accept the suggestions with a selection brush (Figure 2d). The suggestions are continuously updated in real-time according to user inputs.

### 3.2 Workflow Clone

The clone tool is common among interactive content authoring systems. Prior methods mostly clone static content such as images [Pérez et al. 2003], textures [Sun et al. 2013], or geometry [Takayama et al. 2011]. The methods in [Xing et al. 2014] can clone sketching workflows. Our system allows users to clone sculpting

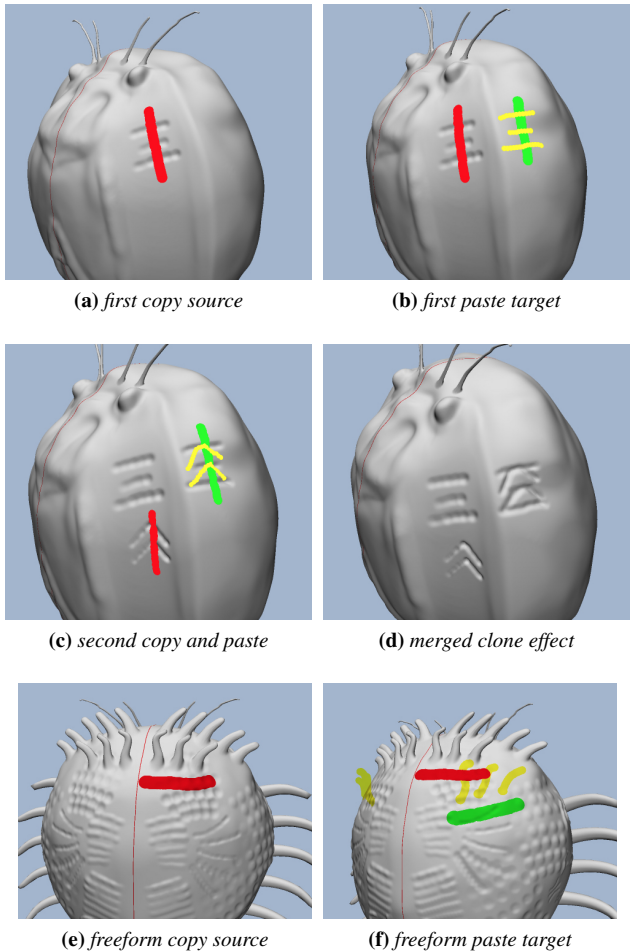

**(a)** *first copy source*    **(b)** *first paste target*

**(c)** *second copy and paste*    **(d)** *merged clone effect*

**(e)** *freeform copy source*    **(f)** *freeform paste target*

**Figure 3:** *Workflow clone example. The red/green brush marks the clone source/target. The clone results are previewed as yellow in (b) and (c) for the users to partially or fully accept. (c) shows clone applied over prior cloned region in (b), (d) shows the merged effect. The clone can be applied to both surface and freeform brushes.*

workflows with more information and options than cloning static geometry. Via our brushing interface, users can select source and target regions, and parameters such as positions, sizes, and directions. Similar to prior clone tools [Kloskowski 2010], our system previews the clone outcomes for which users can accept or ignore. Furthermore, our workflow clone can be applied to already cloned regions, which is difficult to achieve via geometry clone. An example is shown in Figure 3.

### 3.3 Camera Control

3D digital sculpting involves not only brush strokes but also camera manipulations, which can also be quite tedious and repetitive. Fortunately, similar to sculpting operations, camera controls also tend to be predictable [Chen et al. 2014].

We provide a basic camera control mode that automatically moves the viewpoints along with the suggested hints (Section 3.1). This choice is inspired by the observation in [Chen et al. 2014] that users often view the target sculpting regions frontal-parallel, and thus more natural and less disorienting for users than other forms of camera control. Users can quickly press a hotkey to return to

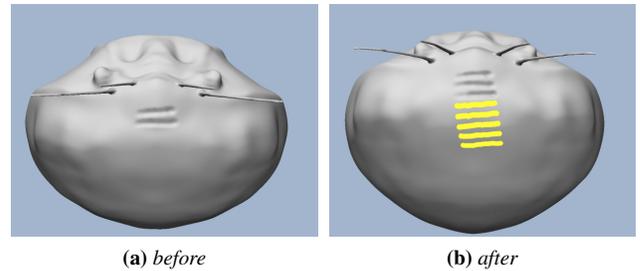

**(a)** *before*    **(b)** *after*

**Figure 4:** *Camera control example. The camera automatically adjusts the viewpoint from (a) according to the suggested brushes in (b).*

the original viewpoint. They can also turn this mode on or off depending on their preferences. One automatic viewpoint example is shown in Figure 4.

### 3.4 Additional Features

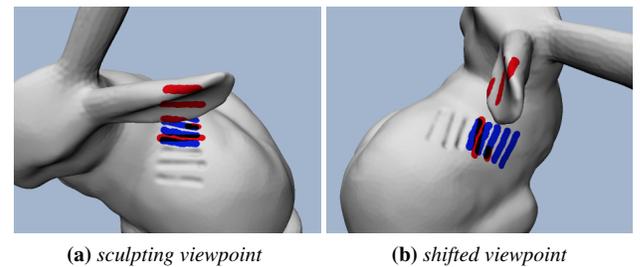

**(a)** *sculpting viewpoint*    **(b)** *shifted viewpoint*

**Figure 5:** *Occlusion example. The blue and red strokes are predictions from the three previous sculpting strokes, with and without considering occlusion. (a) is the original sculpting view, and (b) shifts the view to show the predicted strokes underneath the occlusion.*

**Occlusion** Our system considers geometry occlusion for predictions and suggestions. As exemplified in Figure 5, suggestions predicted on the view-plane (like the canvas for 2D prediction in [Xing et al. 2014; Xing et al. 2015]) may reside on the wrong part of the occluding geometry due to viewpoint, as shown in the red strokes. Our system considers the underlying shape and propagates the predictions more accurately, as shown in the blue strokes.

**Symmetry** Similar to [Calabrese et al. 2016] and existing mainstream sculpting tools, we provide symmetry mode to mirror all brush strokes on the other side of the same model. Users can turn the symmetry mode on and off, and when enabled, the predictions and suggestions will automatically consider symmetry.

**Lock** Locking is an option under some digital sculpting tools (e.g. [Blender Foundation 2016]) to keep some parts of the model fixed while users manipulating other parts. This feature, while useful, can be tedious and difficult for users, especially novices. In addition to future operations, our system can also predict what might need to be locked based on workflows, as exemplified in Figure 6b; such scenarios can be challenging to analyze via geometry only. This locking mechanism can be applied to not only existing geometry as described above but also future geometry predicted by our system which cannot be manually locked, as shown in Figure 6c. Users can easily enable or disable the locking effects via a hot key and thus maintain full control.

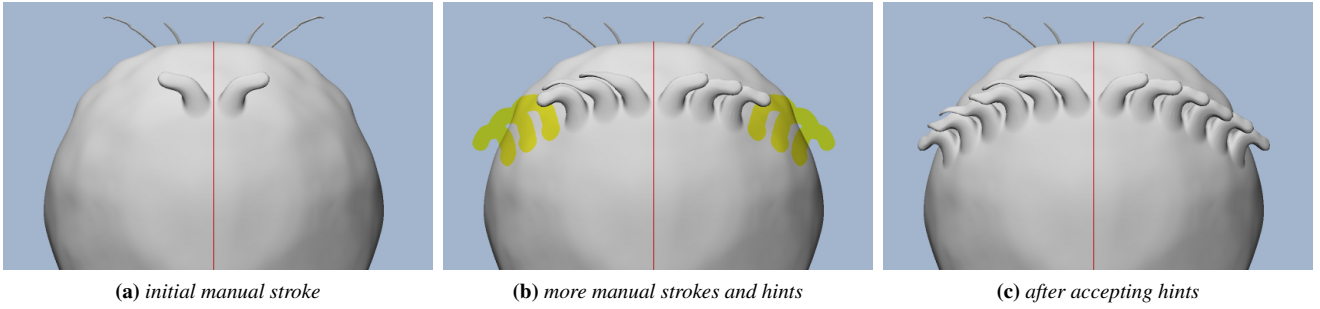

**(a)** *initial manual stroke*  **(b)** *more manual strokes and hints*  **(c)** *after accepting hints*

**Figure 6:** *Workflow lock. After a manual stroke in (a) under the symmetry mode, the user went on to place two more strokes in (b). The yellow parts indicate suggested hints. For comparison, the left side has no workflow lock; notice how earlier strokes can be unintentionally deformed by the later strokes. Our workflow lock can prevent this from happening for both existing geometry and accepted hints, as shown on the right side of (c). Note that the predicted strokes (yellow) are always correct, with or without workflow lock. However, when the user accepted the hints, they will play out in the workflow order as in manual strokes. Thus, without workflow lock, later hint strokes can still deform earlier hint strokes.*

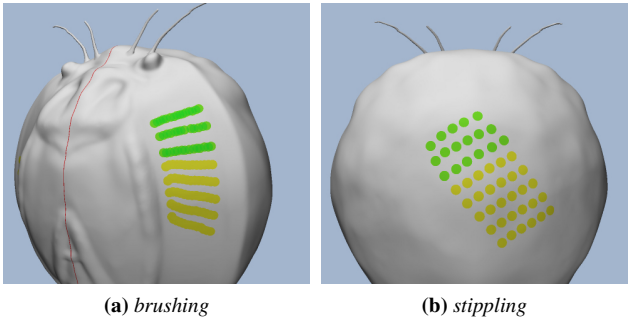

**(a)** *brushing*   **(b)** *stippling*

**Figure 7:** *Autocomplete surface painting examples. Hints are shown in yellow for the users to partially or fully accept.*

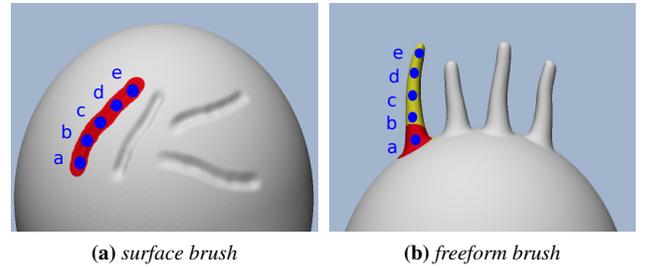

**(a)** *surface brush*   **(b)** *freeform brush*

**Figure 8:** *Brush types. A surface brush (a) has all samples on the object surface, such as the 5 surface samples $s_a$, $s_b$, $s_c$, $s_d$, and $s_e$. A freeform brush (b) has the first sample $s_a$ on the object but the rest for 3D movements such as extrusion $s_a \to s_b \to s_c \to s_d \to s_e$.*

**Painting** After sculpting, users can paint colors over the models, similar to painting in [Pixologic 2011]. Our system can also auto-completes surface painting by extending the 2D autocomplete method in [Xing et al. 2014] over 3D surfaces. An example is shown in Figure 7.

## 4  Method

We describe algorithms behind our autocomplete sculpting user interface in Section 3.

### 4.1  Representation

**Brush** Our system supports two main types of brushes as in common digital sculpting: *surface* brushes for small scale displacements (e.g. clay, crease, smooth), and *freeform* brushes for larger scale shape deformation (e.g. extrusion [Santoni et al. 2016; Denning et al. 2015], drag, grab). They are represented as follows.

**Sample** We represent each brush stroke $\mathbf{b}$ as a collection of point samples $\{s\}$. Each sample $s$ is associated with a set of attributes $\mathbf{u}$:

$$\mathbf{u}(s) = (\mathbf{p}(s), \mathbf{a}(s), \mathbf{t}(s)) \qquad (1)$$

, where $\mathbf{p}(s)$ is the 3D position of $s$, $\mathbf{a}(s)$ is a set of appearance parameters (such as size, type, and pressure) and geometry signatures (such as normal and curvature), $\mathbf{t}(s)$ indicates temporal parameters that include the global time stamp and a sample-id for the relative position within the stroke.

As shown in Figure 8, for a surface brush $\mathbf{b}$, its samples positions $\mathbf{p}(\mathbf{b}) = \{\mathbf{p}(s)\}_{s \in \mathbf{b}}$ all lay on the object surface; while for a freeform brush, $\mathbf{p}(\mathbf{b})$ consists of two parts: the first sample is on the surface, and the rest $(\|\mathbf{p}(\mathbf{b})\| - 1)$ samples are in 3D free space with movement directions controlled by the users.

**Mesh** We adopt a mesh-based representation with two operators, sculpt $\mathbf{c}$ and mesh $\mathbf{m}$, to support geometry and topology changes. A mesh $\mathcal{M}$ is represented by a set of elements including vertices, edges, and faces. Each sculpt operator $\mathbf{c}$ applies specific geometry transformation to mesh elements, such as vertex positions, within a finite support defined by the brush radius. A mesh operator $\mathbf{m}$ can change the underlying mesh resolution and topology by adding or removing mesh elements. The result of each brush stroke over $\mathcal{M}$ is the joint effect of $\mathbf{c}$ with $\mathbf{m}$:

$$\mathcal{M} \leftarrow (\mathbf{c} \otimes \mathbf{m})(\mathcal{M}) \qquad (2)$$

, where $\otimes$ combines $\mathbf{c}$ and $\mathbf{m}$ to achieve Blender-Dyntopo-like or Sculptris-like adaptive tessellation effect, as shown in Figure 9.

### 4.2  Measurement

Analogous to prior methods in predicting 2D sketch [Xing et al. 2014] and animation [Xing et al. 2015], a core component for our method is to measure similarity between 3D brush strokes based on their spatial-temporal neighborhoods. This similarity in turn enables our method to detect repetitions, suggest future edits, clone workflows, and auto-lock brushes. However, unlike [Xing et al. 2014; Xing et al. 2015] where the underlying domain is a fixed 2D plane (drawing canvas), our base domain is a 3D object under dynamic modification. Thus, all definitions of neighborhood and similarity must be conditioned on 3D object surfaces.

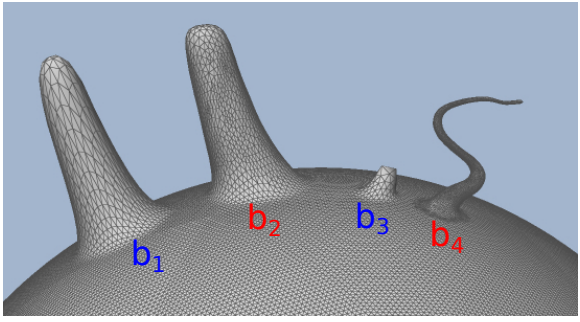

**Figure 9:** *Mesh sculpting effects. A sculpt operator* **c** *such as drag can influence mesh geometry but not topology as shown in* $\mathbf{b}_1$ *and* $\mathbf{b}_3$ *with different brush radii. A mesh operator* **m** *can change mesh resolution and connectivity as shown in* $\mathbf{b}_2$ *and* $\mathbf{b}_4$.

**Neighborhood** We define the neighborhood $\mathbf{n}(s)$ of a sample $s$ as the set of all samples within its spatial-temporal vicinity analogous to the spatial-temporal neighborhoods in [Ma et al. 2013]. Each spatial neighborhood is oriented with respect to a local frame **o** associated with $s$. For surface-brush samples, all spatial distances are computed geodesically (Figure 8a), while for freeform-brush samples via their free space sample distances (Figure 8b). The temporal neighborhood is causal and contains only samples drawn before $s$.

Brush strokes are composed of samples and could capture the high-level relationships between one another. Thus analogous to [Xing et al. 2014; Xing et al. 2015], we use brush strokes as the fundamental units for sculpting workflow analysis and synthesis. The neighborhood of a stroke **b** is defined as the union of its sample neighborhoods:

$$\mathbf{n}(\mathbf{b}) = \bigcup_{s \in \mathbf{b}} \mathbf{n}(s) \qquad (3)$$

**Similarity** For each neighborhood sample $s' \in \mathbf{n}(s)$, we define its differential with respect to $s$ as:

$$\hat{\mathbf{u}}(s', s) = \begin{pmatrix} w_\mathbf{p} \hat{\mathbf{p}}(s', s) \\ w_\mathbf{a} \hat{\mathbf{a}}(s', s) \\ w_\mathbf{t} \hat{\mathbf{t}}(s', s) \end{pmatrix} \qquad (4)$$

, where $\hat{\mathbf{p}}$, $\hat{\mathbf{a}}$, and $\hat{\mathbf{t}}$ represent the sample pair differentials in position **p**, appearance **a**, and temporal parameters **t** defined in Equation (1), and $w_\mathbf{p}$, $w_\mathbf{a}$, $w_\mathbf{t}$ are the corresponding scalar weightings.

We compute the sample position differentials $\hat{\mathbf{p}}(s', s)$ via:

$$\hat{\mathbf{p}}(s', s) = \ddot{\mathbf{p}}(s') - \ddot{\mathbf{p}}(s) \qquad (5)$$

, where $\ddot{\mathbf{p}}(s)$ is the local position of $s$ with frame $\mathbf{o}(s)$ as described in Section 4.3 and relates to the global $\mathbf{p}(s)$ via a coordinate transformation.

From Equation (4), we define the differential between two strokes $\mathbf{b}'$ and $\mathbf{b}$ via their constituent samples:

$$\hat{\mathbf{u}}(\mathbf{b}', \mathbf{b}) = \{\hat{\mathbf{u}}(s', s) | s' = m(s) \in \mathbf{b}', s \in \mathbf{b}\} \qquad (6)$$

, where $m$ is the matching sample computed via the Hungarian algorithm as in [Ma et al. 2013; Xing et al. 2015].

From Equation (6), we can compute the distance between two stroke neighborhoods $\mathbf{n}(\mathbf{b}_o)$ and $\mathbf{n}(\mathbf{b}_i)$ as follows:

$$\|\mathbf{n}(\mathbf{b}_o) - \mathbf{n}(\mathbf{b}_i)\|^2 = \|\hat{\mathbf{u}}(\mathbf{b}_o, c_o) - \hat{\mathbf{u}}(\mathbf{b}_i, c_i)\|^2 + \sum_{\mathbf{b}'_o \in \mathbf{n}(\mathbf{b}_o), \mathbf{b}'_i \in \mathbf{n}(\mathbf{b}_i)} \|\hat{\mathbf{u}}(\mathbf{b}'_o, \mathbf{b}_o) - \hat{\mathbf{u}}(\mathbf{b}'_i, \mathbf{b}_i)\|^2 \qquad (7)$$

, where the first term measures the distance between the two strokes $\mathbf{b}_o$ and $\mathbf{b}_i$ with respect to their central samples $c_o$ and $c_i$:

$$\hat{\mathbf{u}}(\mathbf{b}, c) = \{\hat{\mathbf{u}}(s, c), s \in \mathbf{b}\} \qquad (8)$$

, and the second term computes the distances between their neighborhood strokes with respect to $\mathbf{b}_o$ and $\mathbf{b}_i$. The stroke pairs $\mathbf{b}'_o$ and $\mathbf{b}'_i$ are matched via the Hungarian algorithm as well.

### 4.3 Parameterization

**Surface stroke parameterization** We extend the stroke parameterization method in [Schmidt 2013] for our surface brushes. Each surface brush is parameterized by the surface normal as the z-direction and the stroke path as the y-direction measured by the arc-length $t$. The x-direction is measured by the geodesic distance $d$. We then apply the single-pass forward propagation [Schmidt 2013] to estimate the parametrization for any sample $s$ within distance $r$ of the stroke, as illustrated in Figure 10.

$$\ddot{\mathbf{p}}(s) = \mathcal{P}_s(s) = (t_s, d_s) \qquad (9)$$

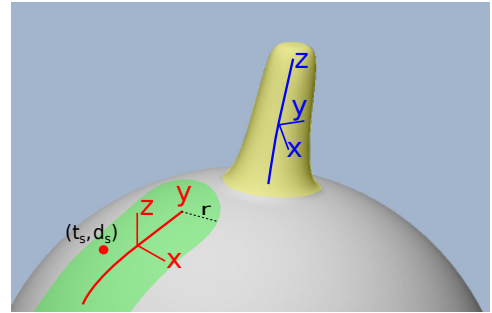

**Figure 10:** *Stroke parameterization. The surface and freeform stroke parameterizations are shown in red and blue with their regions in green and yellow.*

**Freeform stroke parameterization** Unlike the surface brushes, freeform brushes do not adhere to the object surfaces. Thus the method in [Schmidt 2013] cannot directly apply. However, we can extend it into the freeform space as follows. We use the brush path as the z-direction similar to the y-direction for the surface brushes, parameterized by arc-length. The cross-product of the z-direction and the camera look-at direction (non-parallel for sculpting) for each brush sample point forms the y-direction. This is illustrated in Figure 10. Unlike surface stroke parameterization which is 2D, the freeform stroke parameterization is 3D:

$$\ddot{\mathbf{p}}(s) = \mathcal{P}_f(s) = (x_s, y_s, z_s) \qquad (10)$$

## 4.4 Synthesis

In order to synthesize the predictions interactively, we extend the texture optimization methodology [Kwatra et al. 2005; Ma et al. 2013; Xing et al. 2014]. With $I$ as the current sequences of strokes ordered by their time-stamps, we synthesize the next stroke $\mathbf{b}_o$ via the following energy formulation:

$$E(\mathbf{b}_o; I) = \min_{\mathbf{b}_i \in I} |\mathbf{n}(\mathbf{b}_o) - \mathbf{n}(\mathbf{b}_i)|^2 + \Theta(\mathbf{b}_o) \quad (11)$$

, where $\mathbf{b}_i$ represents the corresponding input stroke with similar neighborhood to $\mathbf{b}_o$. The first term evaluates the neighborhood similarity between $\mathbf{b}_i$ and $\mathbf{b}_o$ as explained in Equation (7). The second term denotes optional, applicant-dependent specifications that can be supplied by the users.

Similar to [Xing et al. 2014], for each output, we start with multiple initializations, and select the most suitable one via search and assignment steps. However, instead of a planar canvas that remains invariant for painting, for sculpting the underlying domain is a 3D object under dynamic modification with different brush types as introduced in Section 4.1. Furthermore, all computations need to support interactive responses.

**Initialization** We initialize future strokes based on local similarity with the existing strokes. For the last sculpted stroke $\mathbf{b}'_o$, we identify a candidate set of matching strokes $\{\mathbf{b}'_i\}$. Each $\mathbf{b}'_i$ provides an initialization $\mathbf{b}_{o,i}$ via its next stroke $\mathbf{b}_i$:

$$\hat{\mathbf{u}}(\mathbf{b}_{o,i}, \mathbf{b}'_o) = \hat{\mathbf{u}}(\mathbf{b}_i, \mathbf{b}'_i) \quad (12)$$

Each $\mathbf{b}_{o,i}$ is computed depending on the brush stroke type − surface or freeform, due to their different parameterizations as described in Section 4.3. For surface strokes, Equation (12) is computed on the local surface parameterization. For freeform strokes, Equation (12) is computed by a two-step process: deciding the starting point on the surface, followed by the freeform space movement from the starting point. This is visualized in Figure 11.

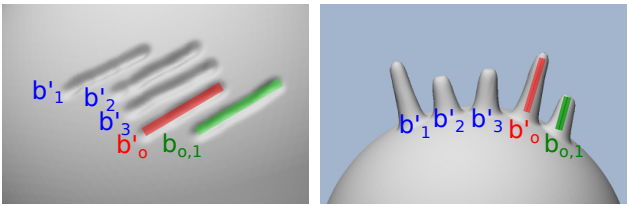

(a) *surface stroke initialization*  (b) *freeform stroke initialization*

**Figure 11:** *Synthesis initialization. For both (a) and (b), the three blue strokes $\mathbf{b}'_{1,2,3}$ are placed in order, before the current stroke $\mathbf{b}'_o$ shown in red. Each of $\mathbf{b}'_{1,2,3}$ can provide a prediction based on its next stroke $\mathbf{b}_1 = \mathbf{b}'_2$, $\mathbf{b}_2 = \mathbf{b}'_3$, $\mathbf{b}_3 = \mathbf{b}'_o$, and $\mathbf{b}'_o$ via Equation (12). For example, the green stroke $\mathbf{b}_{o,1}$ is predicted from $\mathbf{b}'_1$ via $\hat{\mathbf{u}}(\mathbf{b}_{o,1}, \mathbf{b}'_o) = \hat{\mathbf{u}}(\mathbf{b}_1 = \mathbf{b}'_2, \mathbf{b}'_1)$.*

For clarify, we use $\mathbf{b}_o$ as the variable to optimize the choice of $\mathbf{b}_{o,i}$ for different matching $\{\mathbf{b}'_i\}$ of $\mathbf{b}'_o$, by minimizing the energy:

$$E(\mathbf{b}_o) = \sum_{s_o \in \mathbf{b}_o} \sum_{s'_o \in \mathbf{b}'_o} \kappa(s_i, s'_i) \left| \mathbf{u}(s_o) - \mathbf{u}(s'_o) - \hat{\mathbf{u}}(s_i, s'_i) \right|^2$$
$$s_i = m(s_o) \in \mathbf{b}_i, \ s'_i = m(s'_o) \in \mathbf{b}'_i \quad (13)$$

, where $\kappa(s_i, s'_i)$ is a weighting parameter inspired by [Ma et al. 2013] for Gaussian falloff with $\sigma_p$ set to 10:

$$\kappa(s_i, s'_i) = \exp\left(\frac{-|\mathbf{p}(s_i) - \mathbf{p}(s'_i)|^2}{\sigma_p}\right) \quad (14)$$

For each initialization $\mathbf{b}_o$, we optimize it by going through the search and assignment steps below, and the one which has the least energy in Equation (11) would be considered as most suitable and selected to be the predicted stroke.

**Search** During this step, for the initialization $\mathbf{b}_o$ obtained above, within its local spatial-temporal window, we search for the matching stroke $\mathbf{b}_i$ whose neighborhood is similar to $\mathbf{n}(\mathbf{b}_o)$ by measuring the neighborhood similarity in Equation (7). Instead of selecting only one matching stroke, for more robust optimization we search for multiple candidates $\{\mathbf{b}_i\}$ whose neighborhood dissimilarity $|\mathbf{n}(\mathbf{b}_o) - \mathbf{n}(\mathbf{b}_i)|^2$ is lower than $2|\mathbf{n}(\mathbf{b}_o) - \mathbf{n}(\mathbf{b}'')|^2$, where $\mathbf{b}''$ has the lowest dissimilarity value.

For acceleration, similar to [Xing et al. 2014], we perform temporal matching followed by spatial matching instead of matching with the whole temporal-spatial neighborhood. In the first step, we conduct temporal matching to search the candidate matching strokes, from which we use spatial neighborhood for further filtering.

**Assignment** The first term in Equation (11) can be expanded via Equation (7).

The second term, $\Theta$, allows users to configure various parameter settings for various effects such as dot, bumpy, or varying-size strokes as in [Pixologic 2015]. This can be achieve by adding constraints $\mathbf{c}$ for various sample attributes:

$$\Theta(\mathbf{b}_o) = \sum_{s_o \in \mathbf{b}_o} |\mathbf{u}(s_o) - \mathbf{c}(s_o)|^2 \quad (15)$$

We decide the next stroke via minimizing Equation (11) with expansions in Equations (7) and (15).

## 4.5 Deployment

Based on the common framework in Section 4.4, we now describe how to support various modes and options in our system.

**Frame choice** For freeform stroke synthesis, by default, we use local frame for better geometry adaptation. But for certain sculpting tasks, user might prefer global frame to achieve specific effects. Our system thus allows users to switch the frame option. Figure 12 provides an example effect.

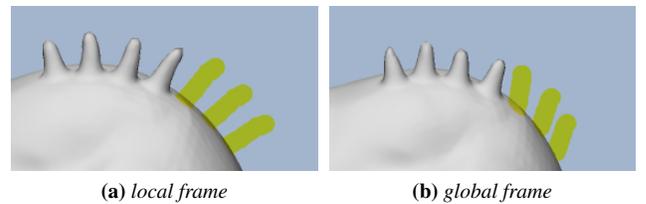

(a) *local frame*  (b) *global frame*

**Figure 12:** *Frame choice example. By default our system synthesizes in local frames as in (a) but users can also opt for the global frames as in (b). The yellow parts are hinted geometry.*

**Hint** The predicted (Section 4.4) and accepted strokes are rendered in light transparent yellow and blue colors to distinguish them from the existing geometry.

**Workflow clone** For workflow clone, we normalize the stroke parameterization to support source and target regions specified with different brush lengths. Specifically, a sample $s$ in a surface stroke would be normalized to be within $t_s \in [0,1], d_s \in [-1,1]$ via:

$$\begin{aligned} t_s &\leftarrow \frac{t_s + r}{T + 2r} \\ d_s &\leftarrow \frac{d_s}{r} \end{aligned} \qquad (16)$$

, where $T$ is the stroke arc length and $r$ is the parameterization width range, as illustrated in Figure 10.

We also normalize the sample-id to fall within $[0,1]$, where 0 and 1 represent the starting and ending positions of brush $\mathbf{b}$.

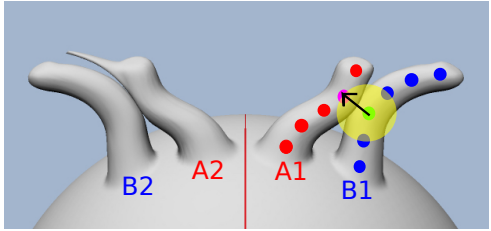

**Figure 13:** *Workflow lock based on spatial-temporal neighborhood. The strokes are placed in the order $\mathbf{b}_A \to \mathbf{b}_B$, and in a symmetry mode for illustration. The left side shows general sculpting effect without workflow lock, the right side is aided with workflow lock. Any samples of $\mathbf{b}_{A1}$ within the spatial-temporal neighborhood of $\mathbf{b}_{B1}$ will be automatically locked, as exemplified in the yellow region of the green sample in $\mathbf{b}_{B1}$.*

**Workflow lock** Automatically deducing which parts of the model to lock based on geometry alone can be challenging, as spatial information may not convey user intention. With workflows, our method can directly associate all brush strokes with the model geometry, and decide what to lock based on workflow similarity as described in Section 4.2. For example, we can lock past workflow samples within a spatial-temporal neighborhood of the current brush stroke, as shown in Figure 13. Based on our experiments, we adopt a simple strategy to lock all past workflows with a spatial-temporal neighborhood of the current brush stroke. This strategy works well when users sculpt in a spatially-temporally coherent fashion, as they often do.

This is also one key difference from [Xing et al. 2014; Xing et al. 2015] where the synthesized sketching strokes are the final outputs. In contrast, sculpting strokes can affect existing geometry.

**Camera control** As described in [Chen et al. 2014; Santoni et al. 2016], user brush strokes tend to correlate with camera movements and thus can facilitate viewpoint selection. Our system stores all camera states, including positions and orientations, as part of the sculpting workflows. Thus, our method is able to predict camera movements in addition to sculpting brushes as described in Section 4.4. In our experiments we have found that excessive camera automation can be disorienting to users. We thus expose only the basic mode of piloting the camera viewpoint along with the predicted next brush strokes.

**Neighborhood and search window** We set $r$ dynamically to be $4\times$ the stroke radius. The spatial-temporal neighborhood of a brush stroke includes its two previous temporal strokes and nearby spatial strokes overlapping its parameterization region (Figure 10). For the search step in Section 4.4, we search within a local temporal-spatial window of 20 previous temporal strokes, and the same spatial neighborhood window as above.

**Neighborhood acceleration** To improve quality and speed, we accelerate the neighborhood matching in Equation (7) by a two-tiered sampling process for the brush strokes. Specifically, we first place three samples uniformly over each stroke to select the most similar candidate strokes, and continue with all samples to select the best matches from the candidates.

**Weights** For Equation (4), we set the position weighting $w_\mathbf{p}$ to be 1. We set $w_\mathbf{a}$ to be 1 if there is no $\Theta$ term in Equation (11), otherwise we set it to be 0.1 and 0.9 for the neighborhood and $\Theta$ terms The $w_\mathbf{t}$ includes global time stamp $w_{\mathbf{t}_1}$ and sample-id $w_{\mathbf{t}_2}$. We set $w_{\mathbf{t}_2}$ to be 1, and $w_{\mathbf{t}_1}$ to be 100 for temporal neighborhood matching to enforce the same sculpting order, and 0 for spatial neighborhood matching.

## 5 User Study

We have conducted a preliminary user study to evaluate the usability and efficacy of our assisted sculpting system. The study considers the following modes: fully manual authoring as in traditional sculpting tools, and our autocomplete functions including hint, workflow clone, camera control, and other features.

**Setup** All tasks were conducted on a 13-in laptop with a Wacom tablet. The study contains three sessions: warm-up, target sculpting, and open creation. The entire study takes about 1.5 hours per participant.

**Participants** Our participants include 1 experienced sculpting modeler and 8 novice users with different levels of sculpting and modeling experiences.

**Warmup session** The warm-up session is designed to help participants get familiar with the sculpting interface and various functions of our system. The tasks consist of adding details via surface brushes, performing extrusions and contour shapes via freeform brushes, and moving the user viewpoints via camera control. One of the authors guided the participants throughout the entire process.

**Target session** The goal is to measure the usability and quality of our assisted system compared to traditional sculpting tools. We asked our collaborating artists to create initial input and reference output (Figure 14). We then asked the participants to start from the initial input and reach the reference output. Each participant performed this task both with and without assisted functions.

**Open creation** The goal of this session is to observe participant behaviors and identity merits and issues of our systems. Participants were free to perform open-ended sculpting using our system with the only requirement of creating structures, either surface or freeform, with at least a certain amount of repetition. One of the authors accompanied the participants through the session and encouraged them to explore different features provided in our system.

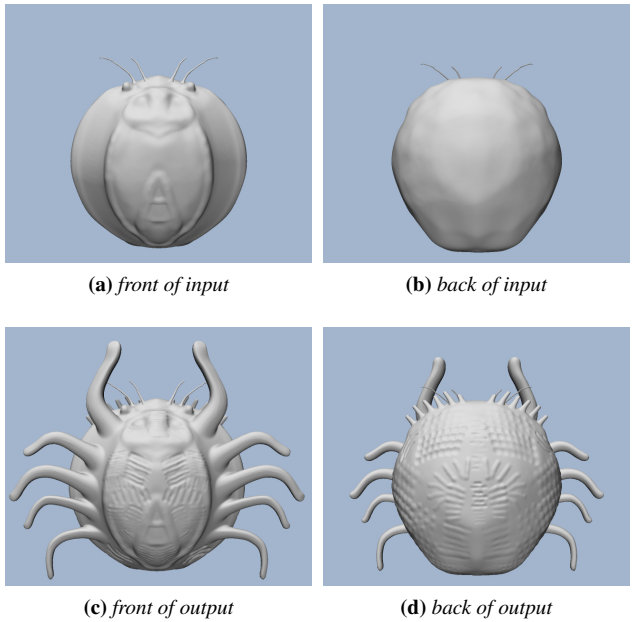

**(a)** *front of input*   **(b)** *back of input*

**(c)** *front of output*   **(d)** *back of output*

**Figure 14:** *Target sculpting task. We provide the base mesh shape shown in (a) and (b), and ask the participants to perform strokes to achieve the desired appearances in (c) and (d).*

## 6 Results

**Objective performance** For the full manual mode, on average the participants took 23 minutes to complete 504 brush strokes. With our autocomplete functions, the participants took 16 minutes to complete 522 brush strokes, including 230 manual and 292 accepted from our hint or clone modes. Thus, our system can help novice users reduce the time and efforts by about 30% and 56% for the particular target study in Figure 14. More detailed statistics are recorded in Appendix A.

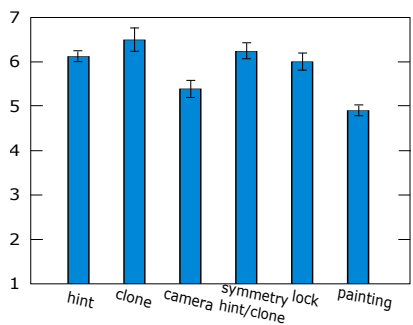

**Figure 15:** *User feedback. Subjective feedback from 8 novice participants in a 7-Likert scale for various modes of our system, expressed as mean ± standard error.*

**Subjective satisfaction** Figure 15 summarizes feedback from the 8 novice participants. Even though some participants had experiences in 2D sketching/painting via digital tablets, we have found it crucial to coach them on practicing how to control the tablet pen more correctly and effectively for 3D sculpting, especially the strength/pressure and camera viewpoint manipulation, as these two factors heavily influence the output visual effects.

Overall, the participants were positive about our system, and gave high ratings to the hint, clone, and lock modes as they are new/interesting ($P_{4,6}$), useful ($P_{2,6,7,8}$), without interrupting usual sculpting process ($P_1$). They preferred the hint mode with brush select for more control. The automatic camera control receives lower rating because it may take time for adjustment and some participants would like better control ($P_1$) and being less conservative ($P_4$). They also provided suggestions for future improvements, such as an eraser function for automatic smoothing ($P_2$), more flexible control for hints (e.g. enlarge the number of suggestions $P_3$), automatic stabilizing brush stroke strength ($P_5$), beautify the UI design ($P_{1,6}$), more powerful prediction for more random strokes ($P_7$). More detailed feedbacks from individual participants are recorded in Appendix A.

**Sample outputs** Figure 16 shows sample outputs from our user study participants. Please refer to our supplementary videos for recorded live actions.

## 7 Limitations and Future Work

We present an autocomplete 3D sculpting system that can reduce input labor and enhance output quality, and demonstrate that the addition of dynamic workflows can effectively augment static geometry for interactive model creation. The autocomplete prototype targets repetitive operations, and falls back to traditional sculpting for non-repetitive operations such as initial shape formation. Assisting the latter via other data-driven or procedural methods in combination with our autocomplete function can be a potential future work.

We propose an automatic camera control mode following workflow suggestions. This is a very basic mode and yet conservative enough to avoid user disorientation. Additional automatic camera controls are possible from the data and method perspectives, but warrant further user studies.

Our current prototype provides basic functionality as a proof of concept. More features in commercial sculpting tools can be added. We also plan to explore other geometry and topology features for more accurate correlation between workflow and shape [Berkiten et al. 2017].

To help prediction, instead of manually crafted algorithms, we are investigating a machine learning approach that analyzes user feedbacks (whether they accept, ignore, or modify the suggestions) for continuous training our prediction model.

Within the scope of this project we have focused on a single user within a single modeling session. The general ideas and specific methods can be extended to multiple users for crowdsourcing and tutorials as well as multiple sessions for individual stylization.

We focus on 3D sculpting as it is a very popular and flexible form of model creation. A potential future work is to consider other forms of 3D modeling, such as VR brushing that operates more directly in 3D free space instead of mostly over the current model surface for sculpting.

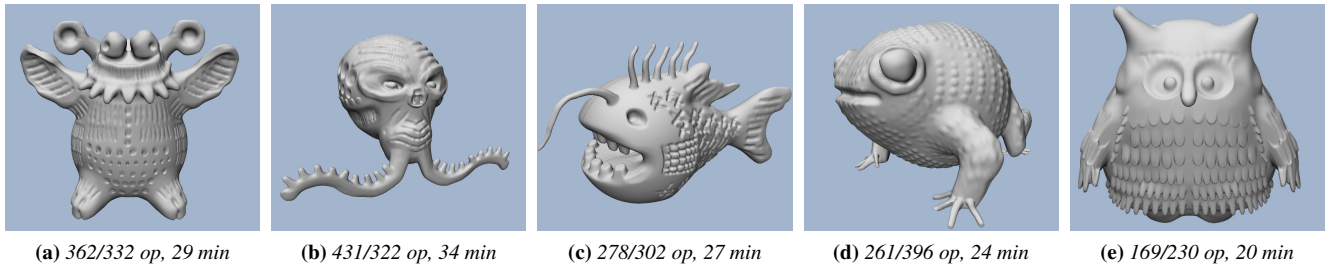

**(a)** *362/332 op, 29 min*    **(b)** *431/322 op, 34 min*    **(c)** *278/302 op, 27 min*    **(d)** *261/396 op, 24 min*    **(e)** *169/230 op, 20 min*

**Figure 16:** *Sample outputs from our participants, all starting from a sphere. Denoted with each output are the following statistics: number of manual sculpting strokes, number of autocomplete strokes, and total authoring time in minutes. Please refer to the supplementary videos for recorded modeling sessions.*

## A  User study

| user | manual | | autocomplete | |
|---|---|---|---|---|
| | min | op | min | op |
| $P_1$ | 22 | 484 | 13 | 212/279 |
| $P_2$ | 24 | 507 | 16 | 229/282 |
| $P_3$ | 26 | 529 | 18 | 240/292 |
| $P_4$ | 29 | 542 | 21 | 262/312 |
| $P_5$ | 25 | 533 | 19 | 225/298 |
| $P_6$ | 18 | 480 | 14 | 217/272 |
| $P_7$ | 18 | 472 | 15 | 220/297 |
| $P_8$ | 20 | 485 | 14 | 235/301 |
| mean | 23 | 504 | 16 | 230/292 |

**Table 1:** *Objective performance from 8 novice participants. Statistics include total completion time in minutes and number of strokes under fully manual mode; total completion time and number of manual/accepted-hint strokes with our autocomplete functions.*

The participant performance has been recorded under Table 1. Below are more detailed participant feedbacks for our user study conducted on January 13, 2017.

Q: What is your score for the following functions, the higher the better?

See Table 2.

| user | hint | | clone | symmetry hint/clone | lock | camera | paint |
|---|---|---|---|---|---|---|---|
| | - select | + select | | | | | |
| $P_1$ | 6 | 7 | 6 | 6 | 6 | 6 | 5 |
| $P_2$ | 6 | 6 | 7 | 6 | 6 | 5 | 5 |
| $P_3$ | 6 | 6 | 5 | 7 | 5 | 6 | 5 |
| $P_4$ | 6 | 6 | 7 | 7 | 7 | 5 | 5 |
| $P_5$ | 6 | 6 | 7 | 6 | 6 | 5 | 5 |
| $P_6$ | 5 | 6 | 6 | 6 | 6 | 5 | 4 |
| $P_7$ | 6 | 6 | 7 | 6 | 6 | 6 | 5 |
| $P_8$ | 5 | 6 | 7 | 6 | 6 | 5 | 5 |
| mean | 5.75 | 6.125 | 6.5 | 6.25 | 6 | 5.375 | 4.875 |

**Table 2:** *Subjective feedback from 8 novice participants in a 7-Likert scale for various modes of our system.*

Q: What are your opinions and suggestions for our prototype, including the main UI and various functions?

$P_1$: It is good that the hints do not interrupt the usual process so I could decide to accept or not, I suggest improving the UI, and some icons to better control the camera position.

$P_2$: Overall, I think it is pretty good and useful software for modelling, but maybe can add an "eraser function":

> After sculpting, I might want to smooth a certain region, but I do not want to use the smooth function again and again. Maybe if you provide a function to let me select the region I want to smooth, and then the whole region could automatically become smooth at one go.

$P_3$: It would be better if the hints for the strokes are more flexible (e.g., the number of suggestions of the potential strokes can be large enough).

$P_4$: The camera movement could be more improved, as current one is a bit conservative. Overall the experience is interesting compared to my usual modeling without hints.

$P_5$: Make users' strength more "stable":

> As sculpting effect has large relationship with the tablet strength, I used tablet before for painting, but for that case, the strength is not that sensitive and important. I wonder whether you could automatically make some "refinements" for those badly controlled strokes, so novices like me could more easily control the stroke.

$P_6$: It is a special experience for me to try this tool, I do a lot modeling via Maya, even though it is very different from yours, the hints and clone are provided as suggestions, and we could preview them. If more functions in other modeling tools could be provided would be better, the UI could be designed to be more beautiful.

$P_7$: If the hints could work for more random strokes, it would be better; the auto and clone functions are useful.

$P_8$: The symmetry mode with hints on both sides helps me reduce manual strokes. I wonder whether painting for such kind of modeling would be useful enough compared to other functions.